\documentstyle[twoside,a4wide]{article}

\input pz.sty
\input epsf.sty

\begin{document}
\PZhead{1}{31}{2011}{7 January}{19 January}

\PZtitletl{GSC 2576--02071 and GSC 2576--01248: two Algol-type
eclipsing binaries}{studied using CCD observations and historical
photographic data}

\PZauth{K.~V.~Sokolovsky$^{1,2}$, S.~V.~Antipin$^{3,4}$, A.~V.~Zharova$^{3}$, S.~A.~Korotkiy$^5$}

\PZinst{Max-Planck-Institute f\"ur Radioastronomie,
Auf dem H\"ugel 69, 53121 Bonn, Germany}
\PZinst{Astro Space Center of Lebedev Physical Institute,
Profsoyuznaya 84/32, 117997 Moscow, Russia}
\PZinst{Sternberg Astronomical Institute,
University Ave. 13, 119992 Moscow, Russia}
\PZinst{Institute of Astronomy,
Russian Academy of Sciences, Pyatnitskaya Str. 48, 119017 Moscow, Russia}
\PZinst{Ka-Dar Public Observatory, Barybino, Domodedovo District,
Moscow Region, Russia}

\PZtyp{ EA }

\PZabstract{ An initial investigation of two poorly studied
eclipsing binaries separated by $\sim 3\arcm$ in the sky is
presented. The first star (GSC 2576--01248) was discovered by the
TrES exoplanet search project. The second one (GSC 2576--02071)
was identified by the authors during CCD observations of GSC
2576--01248. We combine our dedicated CCD photometry with the
archival TrES observations and data from the digitized
photographic plates of the Moscow collection to determine periods
of the two variable stars with high precision. For GSC
2576--01248, addition of historical photographic data provides a
major improvement in accuracy of period determination. No evidence
for period change in these binary systems was found. The
lightcurve of GSC 2576-01248 is characterized by a prominent
variable O'Connell effect suggesting the presence of a dark
starspot and asynchronous rotation of a binary component. GSC
2576--02071 shows a shift of the secondary minimum from the phase
$=0.5$ indicating a significant orbit eccentricity. }

\PZbegintext{K.V. Sokolovsky et al.: GSC 2576--02071 and GSC
2576--01248}

\section{{\large \bf GSC 2576-01248: a short-period Algol-type binary with O'Connell effect}}

GSC~2576--01248 = T-CrB0-04444 = 2MASS~16113251+3306054
(16\hr11\mm32\fsec51 +33\deg06\arcm05\farcs4, J2000; Skrutskie et
al. 2006) was discovered as an eclipsing binary by Devor et al.
(2008) during the Trans-atlantic Exoplanet Survey (TrES; Alonso et
al. 2004). It was also independently detected by the authors using
archival photographic data during initial tests of our digitized
photographic plate analysis pipeline (e.g., Sokolovsky 2006;
Kolesnikova et al. 2010). The digitized photographic plates were
originally obtained in 1973--1975 with the 40-cm astrograph of the
Crimean Laboratory of Sternberg Astronomical Institute (series
``A'' of the Moscow plate collection, field $\xi$~CrB). To assess
the quality of the photographic photometry obtained with our
pipeline, in 2006 we have conducted V-band CCD observations of the
binary using the 50-cm Maksutov telescope (AZT-5) of the Crimean
Laboratory. The CCD observations were reduced in the standard way
involving dark-frame and flat-field corrections and aperture
photometry of the target star, the comparison (TYC 2576--2051--1,
V=10.76, H{\o}g et al. 2000) and check stars in the field of view.
A typical accuracy of individual CCD photometric measurements is
$\sigma_{\mathrm{CCD}}  \approx 0\fmm02$ while the typical
accuracy of our photographic photometry for a $13\mm$--$14\mm$
star was found to be $\sigma_{\mathrm{phot}} \approx 0\fmm07$.

Our series of CCD frames was checked for presence of other
variable objects in the field using the VaST software (Sokolovsky
and Lebedev 2005), which resulted in the discovery of another
eclipsing binary, GSC~2576--02071, just $3\arcm$ away (see the
following section). The effort to determine the period of the
newly discovered binary using observations with a number of
telescopes (Table 1) in 2006--2007 has resulted in a high-quality
lightcurve also for GSC 2576--01248 since both stars fit in the
same field of view for all the instruments used (Fig.~1).

\PZfig{0.7\textwidth}{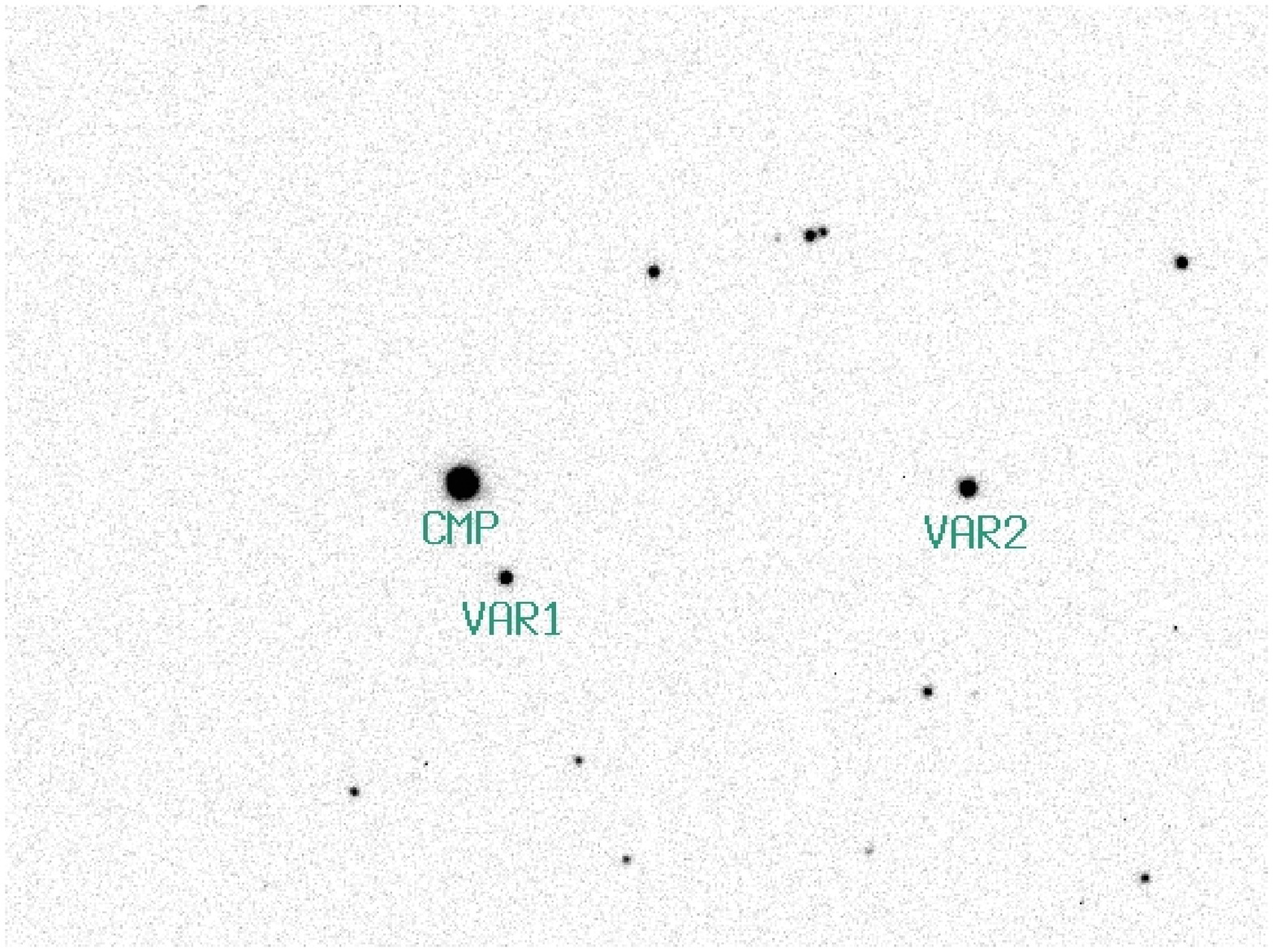}{Finding chart for the
variable stars GSC~2576--01248 (marked VAR1), GSC~2576--02071
(VAR2), and the comparison star TYC~2576--2051--1 (CMP) made from
a $V$-band frame obtained with AZT-5. The field is $8\farcm7
\times 6\farcm5$ wide, centered at 16\hr11\mm28\fsec8,
+33\deg06\arcm55\farcs8 (J2000). North is up, east is to the
left.}

Observations obtained with different instruments were combined
into a single lightcurve by shifting individual lightcurves to
match the average out-of-eclipse brightness to that observed with
AZT-5. The eclipse depth is the same for both $V$-band and
unfiltered lightcurves (Fig.~2) thanks to the spectral response
curve of the blue-sensitive KAI-2020M CCD chip used in the
ST2000XM camera, which reaches its peak quantum efficiency in the
410--550~nm range around the $V$~band.

The combined lightcurve of GSC 2576--01248 was analyzed with the
Lafler \& Kinman (1965) method, which is well-suited for eclipsing
binaries of the Algol type, such as the two variable stars
presented in this paper. Figure~3 shows the fine structure of the
most significant peak on the Lafler \& Kinman periodogram with and
without the archival photographic data included in the analysis.
Thanks to the large time baseline (photographic observations were
obtained in 1973--1975 while the CCD observations were obtained in
2005--2007), the addition of less-precise photographic data to the
high-accuracy CCD observations provides a significant improvement
in the periodogram peak localization resulting in the following
light elements:

$$\begin{array}{r r c r c c}
\mathrm{HJD(TT)_{min~I}} = & 2454255.4056 & + & 0\fday524016 & \times & E. \\
                       &    \pm0.0002 &   &  \pm0.000002 &        &            \\
\end{array}$$

The phased lightcurves folded with the above period are presented
in Fig.~4. The eclipses are partial and last for about 2~hours.
The primary eclipse is $0\fmm66$ deep and the secondary eclipse is
$0\fmm47$ deep in the $V$~band.

Figure~5 shows the phased lightcurve for the two consecutive
observing seasons, 2006 and 2007. The presence of a prominent
($|\Delta m| \approx 0\fmm03$) time-variable O'Connell effect
(O'Connell 1951, Davidge \& Milone 1984) is evident from the plot.
The difference in maxima brightness (and eclipse depth) seems to
be caused by a wide dip, slowly traveling backwards through the
phased lightcurve. The effect may be explained by the presence of
a dark starspot on one of the binary system components that
rotates asynchronously (slightly slower) with respect to the
orbital motion.

Times of individual minima were measured from the CCD lightcurves
by applying the standard Kwee \& van Woerden (1956) method. A
relatively large number of minima observed over the long time
baseline (Table~2) provides an opportunity to search for possible
period changes. From the $O-C$ plot (Figure~6) it is evident that
the long-term behavior of the system is consistent with the
orbital period being the same in 1973--1975 and 2005--2007. The
estimated (also following Kwee \& van Woerden 1956) errors of
minima time determination cannot account for the observed scatter
of measurements around $O{-}C = 0$; however, we note that these
error estimations should be treated with caution since they do not
account for any systematic difference in the shape of the
descending and ascending branches of the eclipse lightcurve (real
or caused by uncorrected systematic effects in photometry) or for
the sensitivity of the method to the exact choice of eclipse
boundaries, which needs to be set manually for the time of minimum
determination. For photographic observations, no time series were
recorded during eclipses. The few faintest data points on the
photographic lightcurve were taken as the best available minima
time estimates in that era. The total uncertainty of the
photographic times of minima determination was set to a typical
exposure time used for the plates ($\sim 40$~min).

\PZfig{0.53\textwidth}{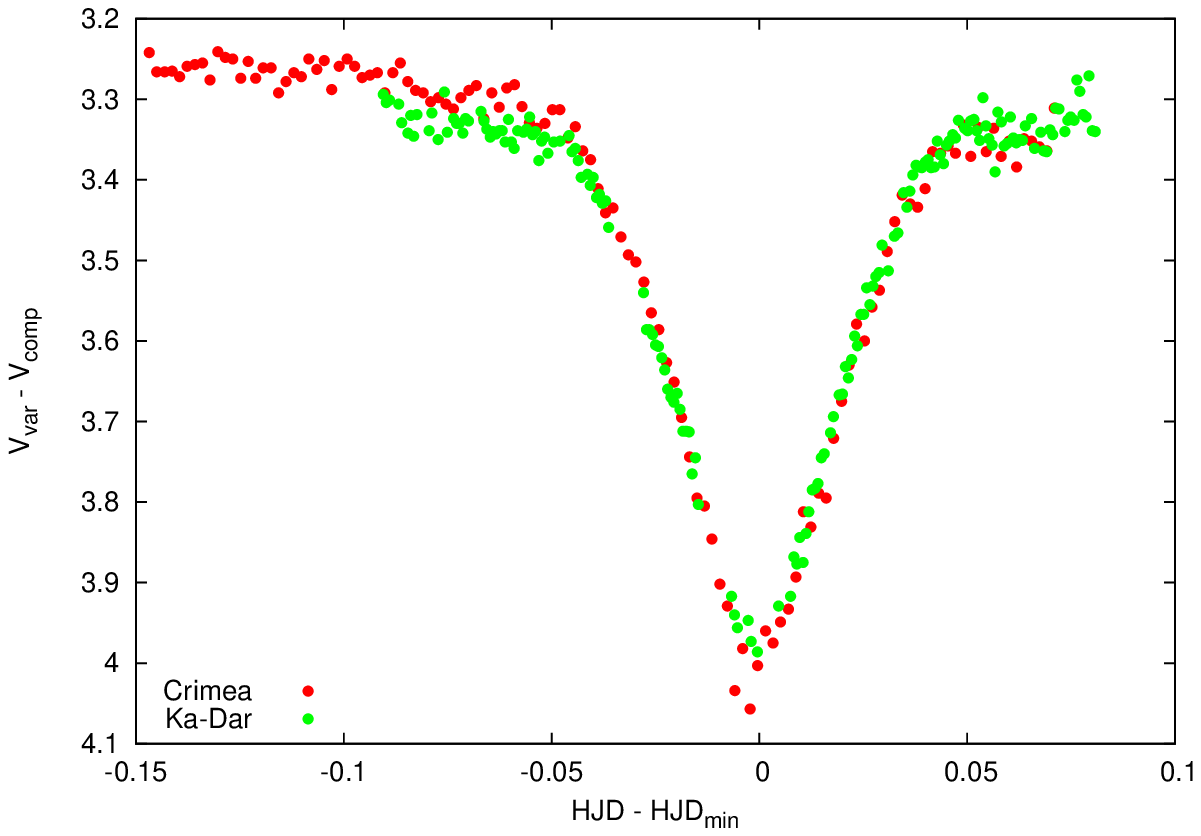}{Two
primary minima of GSC 2576--01248 observed in Crimea (red) with
AZT-5 in $V$~band on JD2453928.4 and in the Ka-Dar observatory
using Meade LX200-14$\arcm$ (green) without filter on JD2454255.3.
The depth of the minima is virtually the same for $V$-band and
unfiltered observations, meaning that by shifting the zero-point
of the unfiltered lightcurve, these observations may be combined
into a single lightcurve for period analysis.}

\PZfig{0.53\textwidth}{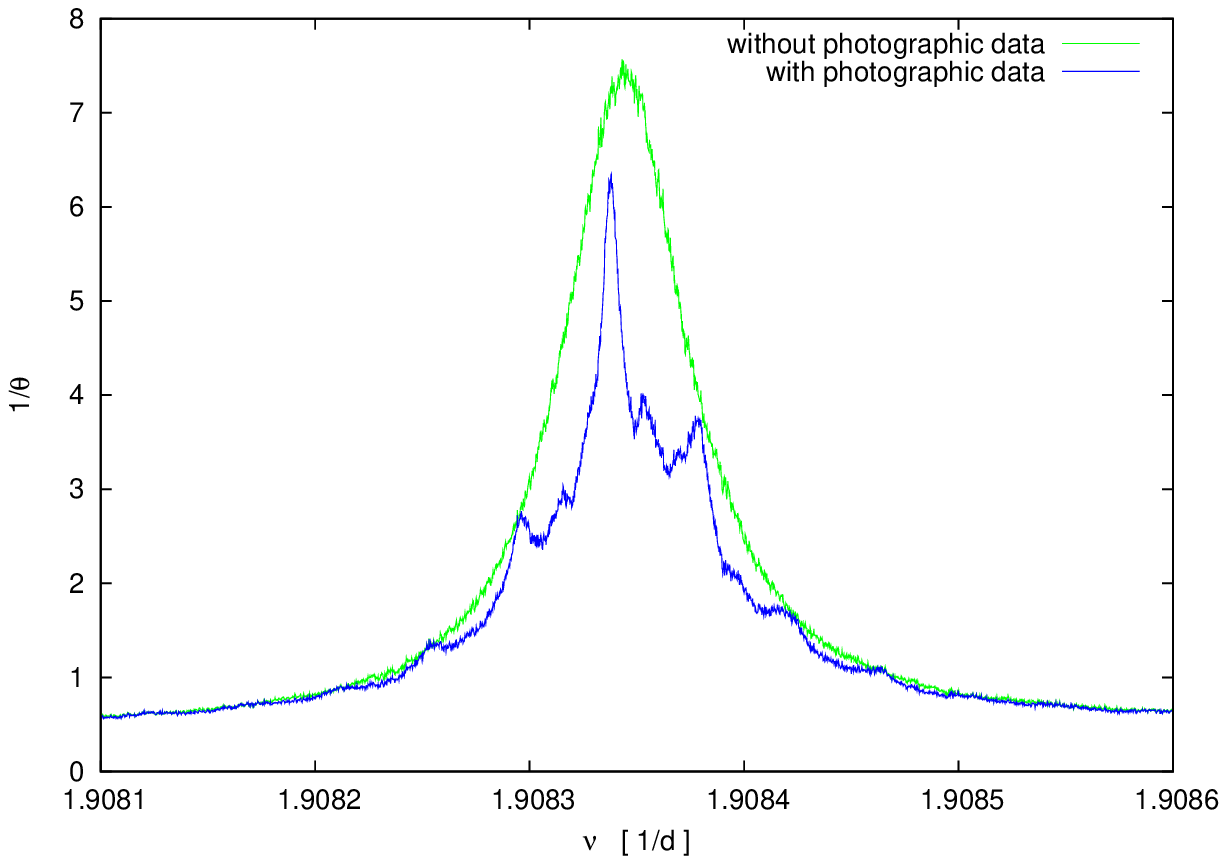}{Periodogram of GSC
2576--01248 computed using the Lafler \& Kinman (1965) method in a
narrow frequency range around the most significant peak. $\nu$ is
the trial frequency in cycles per day; $\theta$ is the normalized
sum of squared magnitude differences between each two subsequent
points of the lightcurve phased with the trial frequency $\nu$
(Goranskij~et~al. 2010). The green curve shows the periodogram for
the CCD-only dataset which includes TrES and our dedicated CCD
photometry. The blue curve represents the dataset that
additionally includes archival photographic measurements,
resulting in a much narrower peak in the periodogram.}

\PZfig{0.53\textwidth}{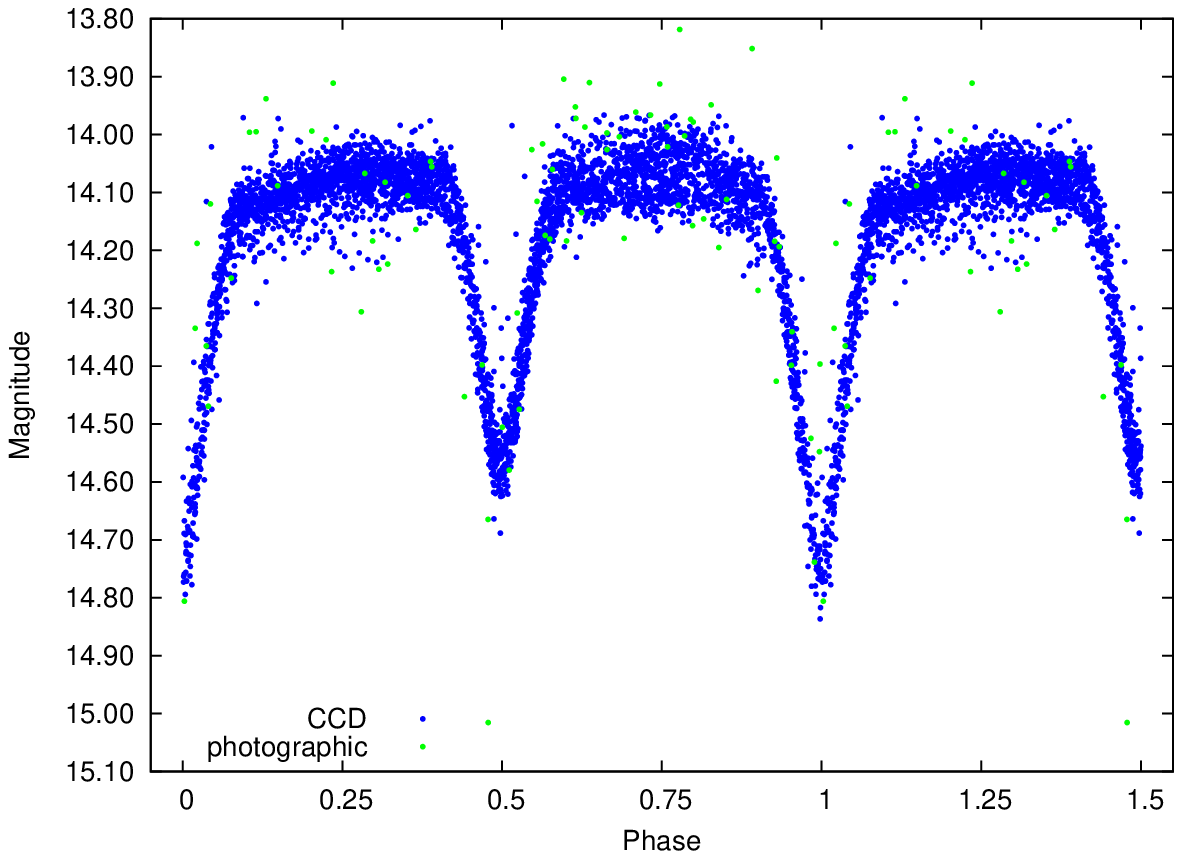}{Lightcurve of GSC
2576--01248 combining CCD (blue) and photographic (green) data
phased with the following light elements:
$\mathrm{HJD(TT)_{min~I}}= 2454255.4056  +  0\fday524016 \times
E$.}

\PZfig{0.53\textwidth}{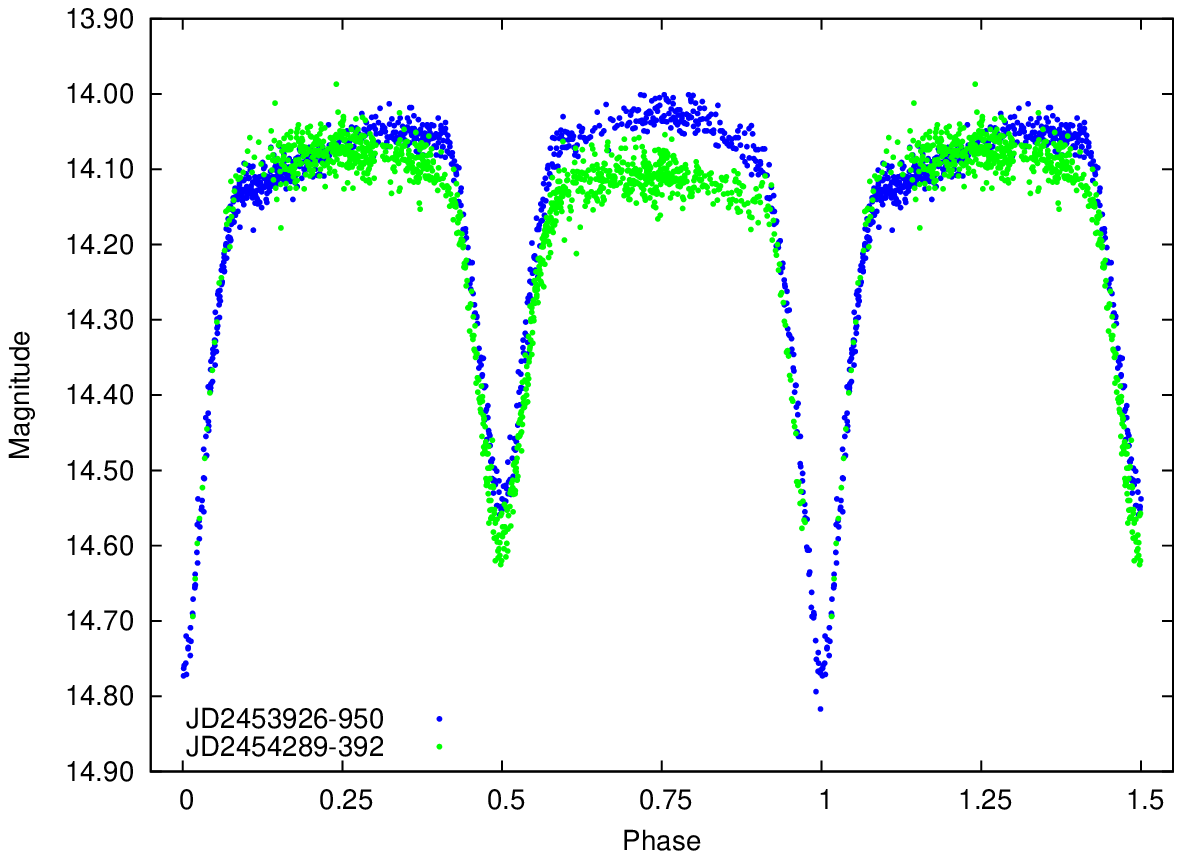}{Lightcurve of
GSC 2576--01248 for the observing seasons 2006 (blue) and 2007
(green) folded with the light elements $\mathrm{HJD(TT)_{min~I}}=
2454255.4056  + 0\fday524016 \times E$. The time-variable
O'Connell effect is clearly visible.}

\begin{table}[b]
{ \centering
 \caption{Instruments used to observe GSC 2576--01248 and GSC~2576--02071}
 \begin{tabular}{lcccl}
 \noalign{\medskip}
 \hline
 \hline
 \noalign{\smallskip}
 Observatory & Telescope & Aperture, cm& Filter & Detector\\
 \noalign{\smallskip}
 \hline
 \noalign{\smallskip}
 Crimea (SAI)      & astrograph& 40    &  none    & photographic plate \\
                   & AZT-5     & 50    &  V     & CCD Pictor 416XTE \\
                   & Zeiss-2   & 60    &  V     & CCD ST2000XM  \\
                   &           &          &  V     & CCD Apogee 47P \\
 Crimea (CrAO)     &           & 24    &  none   & CCD ST2000XM \\
 Ka-Dar            & Meade LX200-14$\arcm$ & 35   &  none    & CCD ST2000XM \\
                   & Vixen     & 10    &  none    & CCD ST2000XM \\
 TrES              & \multicolumn{3}{c}{see Alonso et al. 2004} & CCD \\
 \noalign{\smallskip}
 \hline
 \hline
 \noalign{\medskip}
 \end{tabular}
}

 SAI stands for the Sternberg Astronomical Institute's Crimean Laboratory,
CrAO is the Crimean Astrophysical Observatory (Ukraine).
\end{table}

\clearpage

\begin{table}
 \centering
 \caption{Observed minima times of GSC 2576-01248}
 \begin{tabular}{cccc}
 \noalign{\medskip}
 \hline
 \hline
 \noalign{\smallskip}
  HJD(TT) & Error (d) & Type & Telescope\\
 \noalign{\smallskip}
 \hline
 \noalign{\smallskip}

$2441868.4437$ & $0.0150$ & II &40-cm ($m_{\mathrm pg}=15.56$) \\
$2441925.3015$ & $0.0150$ & I & 40-cm ($m_{\mathrm pg}=15.28$) \\
$2442127.5787$ & $0.0150$ & I & 40-cm ($m_{\mathrm pg}=15.09$) \\
$2442269.3137$ & $0.0150$ & II &40-cm ($m_{\mathrm pg}=15.21$) \\
$2442301.2939$ & $0.0150$ & II &40-cm ($m_{\mathrm pg}=15.12$) \\
$2442487.5805$ & $0.0150$ & I & 40-cm ($m_{\mathrm pg}=15.35$) \\
$2442507.4844$ & $0.0150$ & I & 40-cm ($m_{\mathrm pg}=15.07$) \\
$2453501.8726$ & $0.0006$ & I & TrES \\
$2453502.9180$ & $0.0003$ & I & TrES \\
$2453503.7031$ & $0.0009$ & II & TrES \\
$2453503.9686$ & $0.0004$ & I & TrES \\
$2453505.7983$ & $0.0008$ & II & TrES \\
$2453506.8454$ & $0.0007$ & II & TrES \\
$2453509.7310$ & $0.0010$ & I & TrES \\
$2453510.7770$ & $0.0010$ & I & TrES \\
$2453511.8259$ & $0.0003$ & I & TrES \\
$2453512.8997$ & $0.0078$ & I & TrES \\
$2453515.7563$ & $0.0013$ & II & TrES \\
$2453518.9025$ & $0.0006$ & II & TrES \\
$2453519.9480$ & $0.0008$ & II & TrES \\
$2453520.7357$ & $0.0010$ & I & TrES \\
$2453523.8815$ & $0.0019$ & I & TrES \\
$2453527.8083$ & $0.0011$ & II & TrES \\
$2453528.8539$ & $0.0015$ & II & TrES \\
$2453529.9059$ & $0.0005$ & II & TrES \\
$2453530.6903$ & $0.0002$ & I & TrES \\
$2453532.7891$ & $0.0013$ & I & TrES \\
$2453533.8342$ & $0.0006$ & I & TrES \\
$2453534.8807$ & $0.0006$ & I & TrES \\
$2453535.9306$ & $0.0012$ & I & TrES \\
$2453536.7175$ & $0.0005$ & II & TrES \\
$2453928.4226$ & $0.0004$ & I & AZT-5 \\
$2453932.3507$ & $0.0010$ & II & AZT-5 \\
$2453938.3775$ & $0.0008$ & I & AZT-5 \\
$2453939.4259$ & $0.0004$ & I & AZT-5 \\
$2453944.4031$ & $0.0005$ & II & AZT-5 \\
$2454255.4058$ & $0.0002$ & I & LX200-14\arcm \\
$2454305.4496$ & $0.0004$ & II & Zeiss-2 \\

 \noalign{\smallskip}
 \hline
 \hline
 \end{tabular}
\end{table}

\clearpage

\PZfig{0.53\textwidth}{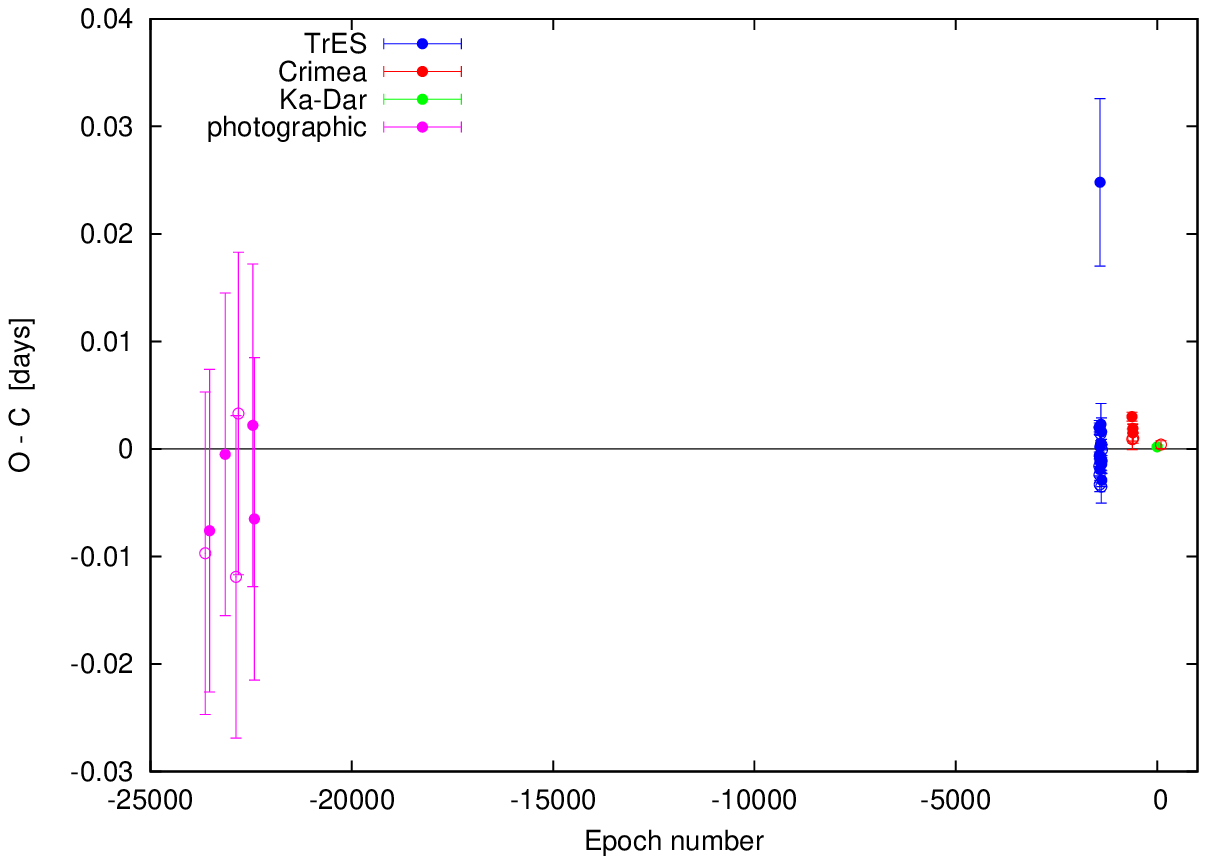}{$O{-}C$ plot for GSC
2576--01248 constructed using the following light elements:
$\mathrm{HJD(TT)_{min~I}}= 2454255.4056  +  0\fday524016 \times
E$. Primary minima are filled circles and secondary minima
(shifted in phase by $0.5$), open circles.}


\section{{\large \bf GSC 2576--02071: a long-period Algol-type binary with an eccentric
orbit}}

GSC 2576--02071 = T-CrB0-03105 = 2MASS~16111795+3307131
(16\hr11\mm17\fsec96 +33\deg07\farcm13\farcs2, J2000; Skrutskie et
al. 2006) was discovered as an Algol-type eclipsing binary during
our CCD observations of GSC 2576--01248. After the initial
detection of an eclipse in summer 2006, it took a significant
amount of observing time with a number of telescopes (Table~1) and
the addition of the TrES archival data to accurately determine the
binary's orbital period:

$$\begin{array}{r r c r c c}
\mathrm{HJD(TT)_{min~I}} = & 2453505.8159 & + & 8\fday28376 & \times & {E}. \\
                           &    \pm0.0007 &   &  \pm0.00025 &        &            \\
\end{array}$$

The archival photographic data provide no additional constrains on
the period because of their low accuracy, small variability
amplitude of GSC 2576--02071, and lacking photographic data around
the primary eclipse. Therefore, only CCD data were used for the
period determination. We have employed the same data analysis and
period search technique as for GSC 2576--01248 (described in the
previous section). The times of minima observed for GSC
2576--02071 are listed in Table~3.

Figure~7 presents the phased lightcurve of GSC 2576--02071. The
secondary eclipse phase is $0.5193 \pm0.0001$, indicating a
significant orbital eccentricity. Additional observations are
desirable to search for signs of an apsidal motion or period
change in this system. The eclipses are partial and last for about
6~hours (Fig.~8). The primary eclipse is $0\fmm35$ deep while the
secondary eclipse is $0\fmm25$ deep. The scatter of the
out-of-eclipse lightcurve (excluding the photographic data) is
characterized by $\sigma = 0\fmm016$, which is consistent with the
expected accuracy of the combined multi-instrument CCD dataset.

\PZfig{0.53\textwidth}{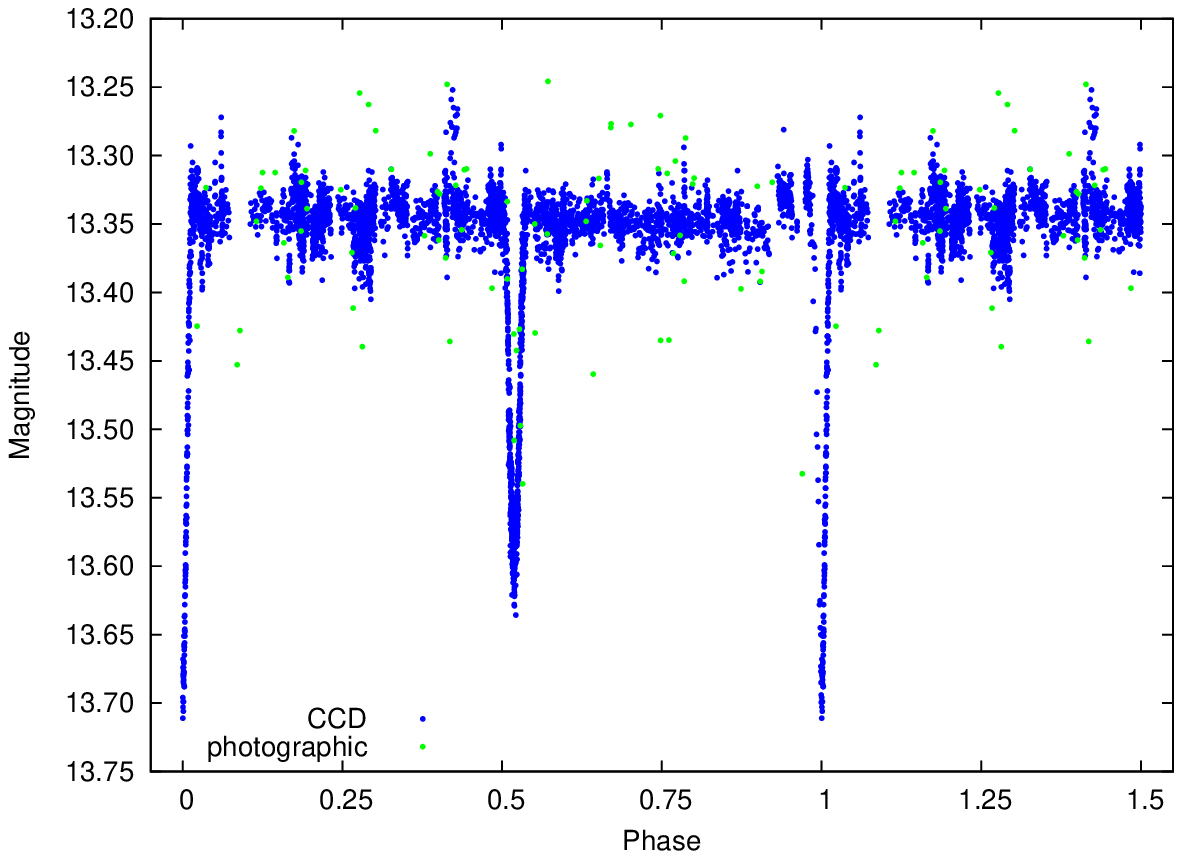}{Lightcurve
of GSC 2576--02071 combining all available CCD (blue) and
photographic (green) photometry phased with the light elements
determined solely from the CCD data: $\mathrm{HJD(TT)_{min~I}}=
2453505.8152  +  8\fday28376 \times E$.}

\PZfig{0.53\textwidth}{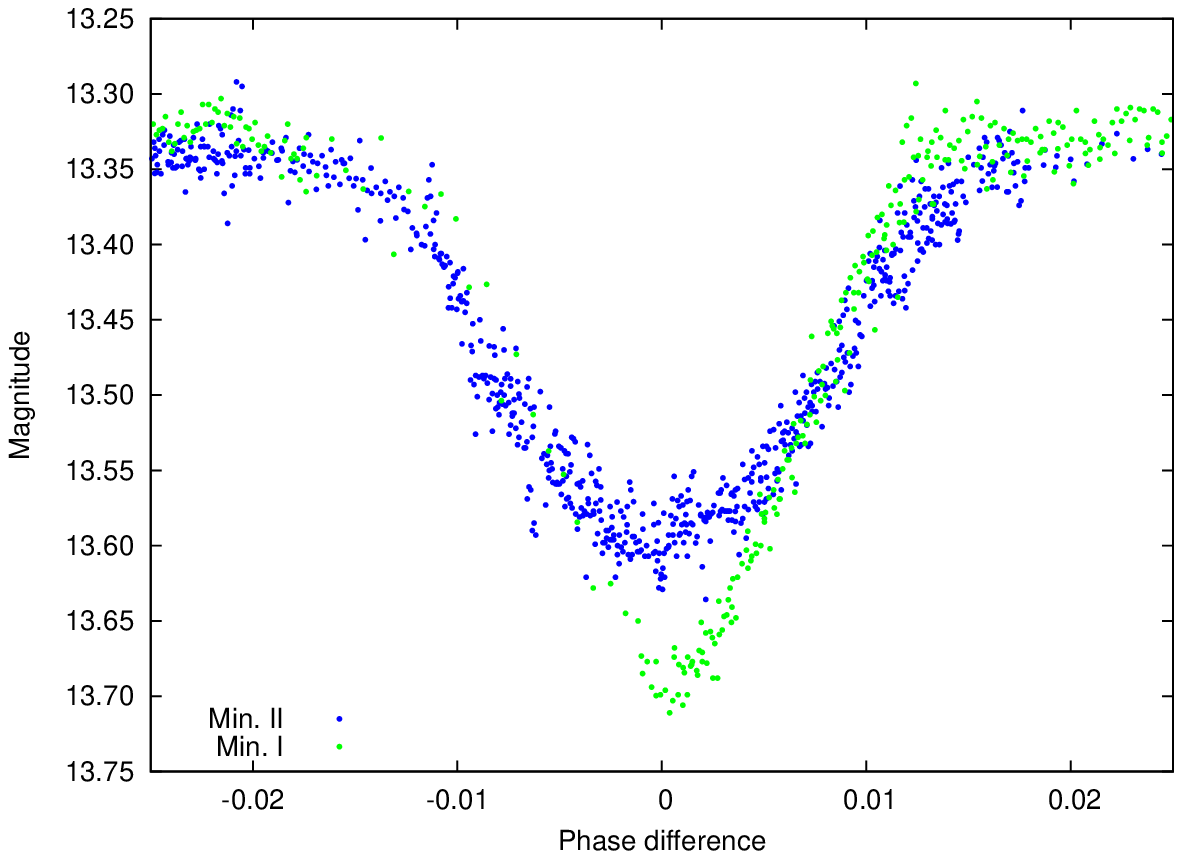}{Primary (green) and secondary
(blue) minima of GSC 2576-02071 compared. The secondary minimum appears to
be about 40~minutes longer then the primary one and is slightly asymmetric
with the ascending branch longer then descending.}

\begin{table}
 \centering
 \caption{Observed minima times of GSC 2576-02071}
 \begin{tabular}{cccc}
 \noalign{\medskip}
 \hline
 \hline
 \noalign{\smallskip}
  HJD(TT) & Error (d) & Type & Telescope\\
 \noalign{\smallskip}
 \hline
 \noalign{\smallskip}
 $2453501.8338$ & $0.0008$ & II &  TrES  \\
 $2453505.8159$ & $0.0007$ & I  &  TrES  \\
 $2454247.3690$ & $0.0007$ & II & LX200-14$\arcm$ \\
 $2454305.3557$ & $0.0007$ & II & Zeiss-2 \\
 \noalign{\smallskip}
 \hline
 \hline
 \end{tabular}
\end{table}

\section{{\large \bf Conclusions}}

An initial investigation of two poorly studied eclipsing binary
stars, GSC~2576--02071 (previously unknown) and GSC~2576--01248
(discovered by Devor et al. 2008), based on the archival
photographic and modern CCD data is presented. In the case of
GSC~2576--01248, the addition of photographic observations from
1973--1975 provides a significant improvement in the accuracy of
period determination over the CCD-only dataset obtained in
2005--2007 thanks to the long time baseline between the
observation epochs. Comparison of the archival photographic data
with modern-day CCD photometry also provides an opportunity to
search for possible long-term period changes, which, however, were
not detected for the two binaries described here. The secondary
minimum of GSC~2576--02071 is shifted from the phase~$=0.5$. This
binary system is a promising apsidal-motion candidate.

{\bf Acknowledgments:} We are deeply grateful to Jonathan Devor,
Francis O'Donovan, Georgi Mandushev, and David Charbonneau of the
TrES team for providing their observations of GSC~2576--02071 and
GSC~2576--01248. We would like to thank S.~V.~Nazarov for the
opportunity to use CrAO's 24-cm telescope and V.~P.~Goranskij for
providing his lightcurve analysis software. This publication makes
use of data products from the Two Micron All Sky Survey, which is
a joint project of the UMass/IPAC-Caltech, funded by the NASA and
the NSF. This work has made use of the Aladin interactive sky
atlas, operated at CDS, Strasbourg, France, the International
Variable Star Index (VSX) operated by the AAVSO, and NASA's
Astrophysics Data System. K.S. is supported by the International
Max-Planck Research School (IMPRS) for Astronomy and Astrophysics
at the universities of Bonn and Cologne.

\references

Alonso, R., Brown, T.~M., Torres, G, et al., 2004, {\it Astrophys.
J.}, {\bf 613}, L153

Davidge, T.~J., Milone, E.~F., 1984, {\it Astrophys. J., Suppl.
Ser.}, {\bf 55}, 571

Devor, J., Charbonneau, D., O'Donovan, F.~T., et al., 2008, {\it
Astron. J.}, {\bf 135}, 850

Goranskij, V.~P., Shugarov, S.~Yu., Zharova, A.~V., Kroll, P.,
Barsukova, E.~A., 2010, {\it Perem. Zvezdy}, {\bf 30}, 4

Kolesnikova, D.~M., Sat, L.~A., Sokolovsky, K.~V., et al., 2010,
{\it Astronomy Reports}, {\bf 54}, 1000

Kwee, K.~K., van Woerden, H., 1956, {\it Bull. of the Astron.
Inst. of the Netherlands}, {\bf 12}, 327

Lafler, J., Kinman, T.~D., 1965, {\it Astrophys. J., Suppl. Ser.},
{\bf 11}, 216

O'Connell, D.~J.~K., 1951, {\it Riverview College Observatory
Publ.}, {\bf 2}, 85

Skrutskie, M.~F., Cutri, R.~M., Stiening, R., et al., 2006, {\it
Astron. J.}, {\bf 131}, 1163

Sokolovsky, K., Lebedev, A., 2005, 12th Young Scientists'
Conference on Astronomy and Space Physics, eds.: Simon, A.,
Golovin, A., Kyiv University Press, p. 79; up-to-date information:
cf. http://scan.sai.msu.ru/vast

Sokolovsky, K.~V., 2006, {\it Perem. Zvezdy Suppl.}, {\bf 6}, 18

\endreferences

\end{document}